\documentclass[runningheads]{llncs}

\usepackage{lmodern} 
\usepackage[T1]{fontenc} 
\usepackage{textcomp} 
\usepackage{graphicx}
\usepackage{cite}
\usepackage{amsmath,amssymb,amsfonts}
\usepackage{algorithmic}
\usepackage{graphicx}

\usepackage{tikz}
\usepackage{balance, hyperref, cleveref}
\usepackage{flushend}
\usepackage{orcidlink,booktabs}

\usepackage{amsmath,amssymb,amstext} 

\usepackage[ruled,linesnumbered]{algorithm2e}
\usepackage{listings, xcolor}

\usepackage{xcolor}
\def\BibTeX{{\rm B\kern-.05em{\sc i\kern-.025em b}\kern-.08em
    T\kern-.1667em\lower.7ex\hbox{E}\kern-.125emX}}
    
\newcommand{\toolad}{AdChain}

\begin{document}
\title{AdChain: Decentralized Header Bidding}
\titlerunning{Decentralized Header Bidding}

\author{Behkish Nassirzadeh \inst{1}\orcidlink{0000-0002-3227-1409} \and
Albert Heinle \inst{2} \and
Stefanos Leonardos \inst{3} \orcidlink{0000-0002-1498-1490} \and 
Anwar Hasan \inst{1}\orcidlink{0000-0003-4103-7945} \and
Vijay Ganesh\inst{4}\orcidlink{0000-0002-6029-2047}
}

\authorrunning{B. Nassirzadeh et al.}
%
\institute{University of Waterloo, Waterloo, Canada \email{\{bnassirz,ahasan\}@uwaterloo.ca} \and 
CoGaurd, Waterloo, Canada \email{albert.heinle@googlemail.com} \and
King's College London, London, UK \email{stefanos.leonardos@kcl.ac.uk} \and 
Georgia Institute of Technology, Atlanta, USA \email{vganesh@gatech.edu}
} 
\maketitle    

\begin{abstract}
Due to the involvement of multiple intermediaries without trusted intermediaries, lack of proper regulations, and a complicated supply chain, ad impression discrepancy plagues online advertising. This issue accounts for up to \$82B of annual revenue loss for the honest parties. This loss can be significantly reduced if there is a precise and trusted decentralized mechanism. This paper presents \toolad, a decentralized, distributed, and verifiable solution that detects and minimizes online advertisement impression discrepancy rate. \toolad\ aims to establish trust by acquiring multiple independent agents that receive and record log-level data and a consensus protocol that determines the validity of each ad data. \toolad\ is scalable, efficient, and compatible with the current infrastructure. Our experimental evaluation on over half a million ad data points discovers systems parameters to achieve an accuracy of 98\% to decrease the ad discrepancy rate from up to 20\% to 2\%. Our cost analysis shows that, on average, active nodes on \toolad \ can generate a profit comparable to miners on main Blockchain networks like Bitcoin.

\end{abstract}
\section{Introduction}
\label{intro}

In 2022, the online advertising industry generated \$567B in worldwide revenue, with a projected growth of over \$836B by 2024 \cite{hpaper}\cite{MarketSize}. In its simplest form, it involves attaching an invisible pixel with an executable JavaScript to an ad placeholder. This pixel returns user interactions, such as views, to a server. Advertisers (ad buyers) pay for an ad only when it results in a view, click, or any other action deemed an {\it impression}. Among various online advertising techniques, {\it header bidding} is the dominant one, accounting for more than 73\% of the market share (\$414B in 2022) \cite{HeaderBidding}.

Traditionally, publishers (website owners) would rely on a single ad server to fill their ad slots. In header bidding, multiple advertisers can bid on the publisher's ad space in real time before a web page loads. Once the winning bid is determined, the chosen ad is delivered to the website visitor when the page loads. This auction system increases competition among advertisers, which increases revenue for publishers \cite{HeaderBidding}. However, one of the complications of this method is that it involves various middlemen, with Supply-Side Platforms (SSPs) and Demand-Side Platforms (DSPs) being the key players. These middlemen independently collect ad data using their own methods and standards.

\begin{figure}[t]
\centerline{\includegraphics[width= 0.7 \textwidth]{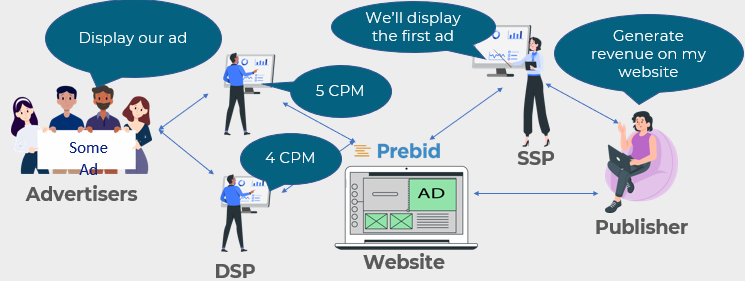}}
\centering
\caption{A Simple View of Header Bidding. CPM is Cost Per Mille}
\label{ad}
\end{figure}

A publisher may not know all the technical aspects of running a proper header bidding action on their site, which is why they hire SSPs. Although SSPs provide a report to the publishers, these reports are generally complicated and too technical; hence, publishers need to trust the SSPs. On the other hand, advertisers may not be able to run ad campaigns or find the right audience and target for their ads. As a result, they rely on DSPs. Therefore, most advertisers do not know where their ads are being shown and must trust the DSPs. The DSPs generally send their bids to SSPs through platforms like Prebid \cite{prebid}. A brief and straightforward overview of header bidding is shown in Figure ~\ref{ad}. It is important to note that there are many other middlemen involved that are not shown here. 

\subsection{Problem Statement}

Ad fraud is the main problem of header bidding. Ad impression discrepancy fraud, or ad discrepancy for short, is the primary type of ad fraud \cite{AdFraud}, \cite{fraud}. Ad discrepancy is the difference between the impression records of the involved parties. Due to the complexity of this method, the regulations by the Interactive Advertising Bureau (IAB) allow for up to 10\% margin of error. If the discrepancy between the SSPs and DSPs is less than 10\%, they have to agree amongst themselves. Because of this allowed margin of error and the fact that there are no set industry standards, this state of affairs offers an excellent possibility for fraudulent players to take advantage of the system by reporting false data to publishers, advertisers, and other involved parties. The discrepancies can even be an issue when a trusted player acts as both the SSP and DSP. This was apparent in the recent case where Facebook admitted that they used the wrong metrics to collect impression data, costing involved parties millions of dollars \cite{Facebook} \cite{fb}.

This problem is mainly addressed by ad-hoc, and often broken, in-house solutions that have not been able to solve the issue reliably \cite{AdFraud}. Although IAB only allows for a 10\% error margin, in reality, even a 20\% discrepancy is considered normal. This generally results in up to 20\% (about \$82 billion) revenue loss due to ad impression discrepancies~\cite{Discrep}, \cite{20p},\cite{Adex}, \cite{Adpushup}. As a result, there is a great incentive to cheat the system and do so repeatedly (i.e., one DSP's number is always lowered by 7\%).

Additionally, in the existing ecosystem, honest players face difficulties surviving. Dishonest players can manipulate their bids and statistics, gaining an unfair advantage. For instance, dishonest players can offer higher bids in the auction, knowing they will manipulate their impression data to pay less in the end. This forces honest parties to gain lower profits or resort to dishonest tactics to compete. As a result, honesty is significantly disadvantaged in this setup. 

In summary, the ad discrepancy problem can be characterized as a collection of interrelated problems. First, advertisers cannot monitor their ad transactions in detail and often do not know where much of their money is going. Then, incompatible formats for ad impression data are being used. This makes it very expensive to discover the reason for ad discrepancies. Consequently, potential fraudulent activity goes undetected, resulting in a revenue loss for advertisers and publishers (and possibly for other middlemen). Finally, this market has no self-correcting mechanism, resulting in honest players being driven out of it.

All these challenges boil down to one major problem: \noindent{\bf  A worldwide online advertisement ecosystem that lacks a transparent, verifiable, and trusted ground truth.} If we had a mechanism to establish such ground truth, we could expose dishonest players and fraudulent activity, resulting in lower costs for all parties involved. Further, any mechanism that enables one to establish verifiable truths is likely to be self-corrective, i.e., encouraging all parties to act with greater honesty and transparency, resulting in fair competition. 

\subsection{Brief Overview of the Solution}
When there is a problem of trust involving multiple parties, one solution that comes to mind is to use Blockchain and Decentralized Oracle Networks (DONs) to access off-chain data. However, online advertising requires a precise counting system, as the revenue is based on the reported count. The existing DONs are ill-suited for counting systems. This is because in the current DONs, some level of error is tolerated, and the results are not precise. Moreover, the current Blockchains cannot scale to the millions of ad TPS. Additionally, the immutability of Blockchains is a drawback in this case. Each ad data needs to be deleted after its short lifespan to save storage. If this data were to be stored permanently, it would require colossal storage. Also, in some cases, the stored data might need to be modified. However, a decentralized oracle counting system like CountChain \cite{CountChain} can be utilized to design a decentralized online ad ecosystem to circumvent the constraints associated with current Blockchain technology effectively. 

In this paper, we present \toolad, a decentralized oracle network that detects and prevents online advertisement fraud. Since it is decentralized and transparent, it is more reliable than the existing solutions. Our solution aims to establish the ground truth and let the market act as a self-corrective mechanism by creating fair competition between involved middlemen. This solution can spot unfair and dishonest middlemen and thus help parties who are not getting their due to take appropriate corrective actions. 

In this design, the consensus protocol for CountChain \cite{CountChain} is tailored for the case of header bidding. Therefore, each ad holder contains multiple invisible pixels. The endpoint of each pixel is randomly assigned to an oracle node during the header bidding auction. These nodes become the verifiers of that ad. When an ad is served, these assigned verifiers receive the ad's log-level data through the invisible pixels. Verifiers pay a fee to submit a boolean proposition to the system upon detecting a valid ad transaction. These propositions undergo a verification process by other verifiers through a consensus protocol. In this protocol, verifiers vote true or false, accompany their votes with proof, and stake a fee. The requirement for monetary staking helps mitigate the risk of Sybil attacks~\cite{sybill}.

Nodes earn points each time they submit a valid proposition or True vote. These points are subsequently transformed into monetary rewards. In the event of discrepancies among multiple parties, the verification process's outcome serves as proof since various random verifiers validate it. Additionally, the received log-level data can be used to identify the source of the discrepancy.

In this paper, we make three major contributions. First is the design and implementation of \toolad, the first decentralized header bidding system to mitigate ad impression discrepancy. Second is an experimental evaluation using more than half a million simulated ad data points to identify specific system parameters that achieve high scalability and up to 98\% accuracy in resolving the discrepancy issue by as much as 90\%, as well as reducing the chance of Sybil attacks. Finally, a cost analysis of the nodes on \toolad \ shows that, on average,  active nodes on \toolad can generate an annual profit of 17,000 USD, which is comparable to the average profit of Bitcoin miners.

\section{Related Work}
\label{related}
Adhash \cite{adHash} is a blockchain that aims to reform the online advertising industry by removing middlemen completely. Alkimi Exchange \cite{alkimi} is a decentralized ad exchange to build a better user experience by replacing the inefficient legacy programmatic ad exchanges with the mission to restore the value exchange between advertisers, publishers, and users. Adwatch \cite{adwatch} is a Blockchain for advertisers to define the rules for each ad individually and track ad campaigns. 

However, these networks require a complete change in the current infrastructure, aim at removing useful and powerful middlemen, or may have some privacy concerns since all data is shared publicly. To the best of our knowledge, \toolad \ is the first decentralized platform that solves the ad discrepancy issue and provides valuable information to find the root cause of the discrepancy. At the same time, it is compatible with the current infrastructure and does not require significant modifications to the current setup.

\section{Preliminaries}
\label{background}

\subsection{Header Bidding}
Publishers can earn money using ad exchanges like AdX (by Google). However, the revenue that can be made is relatively minor compared to an open auction with other players. This auction mechanism is usually called header bidding, which transpires through platforms such as Prebid \cite{prebid}. Because of the complications of this method, SSPs usually handle these auctions from the publisher's side. Upon initiating a website loading event by a visitor in this auction, multiple DSPs engage in bidding, which is contingent on user quality and specific criteria. The winning bid is determined almost instantly, and the chosen ad is delivered to the visitor when the page loads. Subsequently, the DSPs and SSPs count the impressions and compile ad-related data. Following a predefined number of transactions or a temporal interval, the DSP remits payment to the SSP, which, after deducting its share, disburses the remainder to the publishers \cite{HeaderBidding}.

\subsection{Prebid}
One of the major tools performing these header bidding auctions is Prebid. Prebid helps put the whole auction into one system \cite{prebid}. At a high level, Prebid steps are as follows:
\begin{enumerate}

    \item The ad server’s tag on the page is paused, bound by a timer, while the Prebid.js library fetches bids and creatives from various SSPs and exchanges.
    \item Prebid.js sends the information about those bids to the ad server’s tag on the page. This information contains cost and action type.
    \item The ad server contains line items targeting those bid parameters.
    \item If the ad server decides Prebid is the winner, the ad server sends a signal to Prebid.js requesting the library to write the winning creative to the page.
\end{enumerate}

\subsection{Our Survey of Major SSPs and DSPs}
We interviewed multiple major SSPs and DSPs. 80\% of them regularly face the discrepancy problem and seek a reliable solution. They mostly use a third party to cross-verify their records with the others to resolve the discrepancy. This approach can restore some of the lost revenue and help the parties find the root cause of the discrepancy in the future. However, it costs millions of dollars annually, often takes 2 to 3 months, and requires a new trusted party. Most of the participants have considered or tried a Blockchain approach. However, they faced precision or scalability limitations. Meanwhile, 10\% of the participants would only consider a new solution if it is inexpensive, as the discrepancies only affect their revenue by less than 5\%. Finally, 10\% of the participants are not significantly affected by the discrepancies.

\begin{table*}[!t]
\small
\newcommand{\tabincell}[2]{\begin{tabular}{@{}#1@{}}#2\end{tabular}}

\caption{Summary of the terminology.}
\label{table0}
\resizebox{\textwidth}{!}{
\begin{tabular}{@{}l@{\hskip 5pt}l@{\hskip 5pt}l@{}}
\toprule
 \textbf{Term} & \textbf{Definition} & \textbf{Set By}\\

\midrule
Ad\_ID &  Unique identifier of an ad & Invisible Pixels  \\
Host & The platform storing all the information & Owner/ Network\\
User\_ID & Unique identifier of the website visitor & Invisible Pixels  \\
Owner & The party interested in obtaining the result of the counting system  & Network \\
Player & Participant nodes that intend to validate ad data& Network/ Nodes \\
Player\_ID & Unique identifier of each Player & Host\\
Player Point & The point accumulated
for each vote or proposition submission & Host \\
prize pool & The financial reward and incentive & Owner and Players \\
Proposition & A Boolean request containing ad data to be validated & Submitter \\
Proposition Countdown & Minimum number of propositions a player need to verify to end the game  & Owner \\
Proposition Deadline & The deadline to vote for a Proposition & Owner/ Host \\
Propositions Pool & The list of available Propositions to be verified & Host \\
Proposition Price & The stake used when the Players vote for the Propositions & Owner \\
Submitter & The Player that submits a Proposition to
be verified & Players and Host \\
Verifier & Randomly chosen Players to verify a Proposition & Host \\
\bottomrule

\end{tabular}
}
\end{table*}

\section{Description of the Design of \toolad}
\label{components}

\subsection{Setup and Terminology}
The system is a decentralized oracle network based on CountChain \cite{CountChain} with some modifications to make it applicable to the case of online advertisement. The terminology summary is provided in Table~\ref{table0}. A permissioned distributed server acts as an independent middleman in this model. Implementing this can be done using a smart contract or a trusted party. We refer to this middleman as the \emph{host}. The host is responsible for storing important information and needs to be able to generate random (pseudo-random) numbers, compute cryptographic hash like SHA3, and send and receive bulk data. Like the other middleman, this mediator also collects ad data using invisible pixels. The entity that runs the host is the \emph{owner}, responsible for ensuring the network operates as expected and defining the criteria for the ad data and other system parameters.

Each ad and website visitor contains a unique identifier, known as \emph{ad\_ID} and \emph{user\_ID}, respectively. The system is based on a set of \emph{propositions} with convert truth values and associated monetary values, represented by \emph{proposition price}, used in a voting game. Each proposition also contains a deadline by which the proposition is decided, the game outcome is computed, and the corresponding rewards and penalties are administered. This is the \emph{proposition deadline}, which is set by the owner. Also, the system’s fund or the \emph{prize pool} is the accumulation of the initial fund and the stakes collected from the dishonest players. The publishers provide the initial fund.

\begin{figure}[t]
\centerline{\includegraphics[width=  \textwidth]{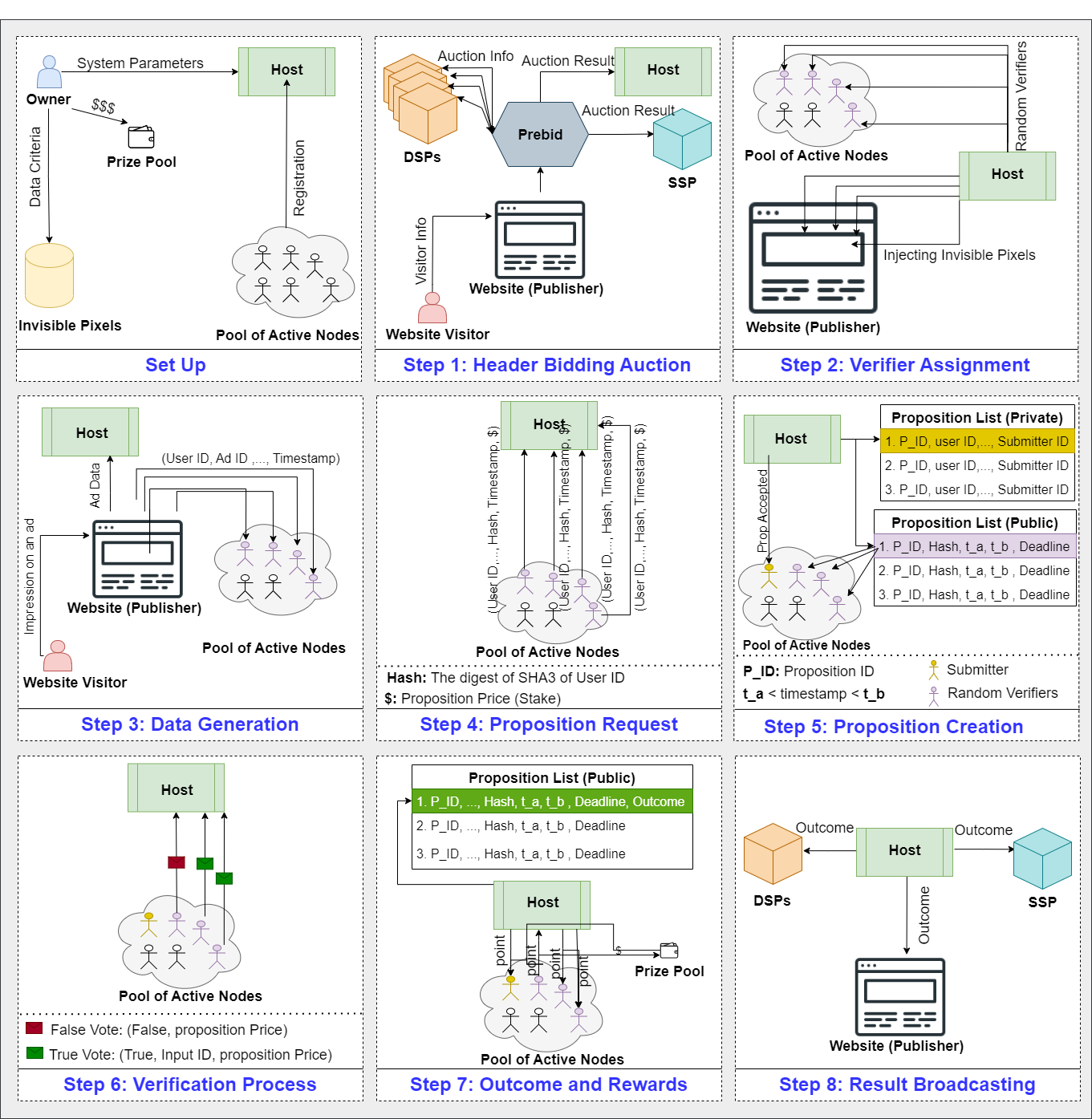}}
\centering
\caption{An Overview of \toolad}
\label{fig2}
\end{figure}

Each node or \emph{player} has two entities of \emph{player\_ID}, a unique identification, and \emph{player point}, the point accumulated for each vote or proposition submission. The financial reward for each player is decided based on the points accumulated by each player. Each player also has another property called \emph{ proposition countdown (PCD)}, which is the number of propositions a player needs to verify before leaving the game and receiving financial rewards. This property is because this system requires an ongoing counter and never ends, as the ad transactions do not stop at any point. Therefore, there has to be a point where players can stop playing if needed and receive financial rewards. Any middleman like SSPs, DSPs, publishers, and advertisers can be a node in this system. Other entities can also sign up and become a node.

Multiple invisible pixels get injected into each ad, and each invisible pixel's endpoint is a random player the host picks. These players become the \emph{verifiers} of that ad transaction and receive data from the pixel independently. Each verifier can submit a proposition request to the host once a valid transaction is detected. The first verifier that submits a valid proposition request becomes the \emph{submitter} of the proposition. Therefore, each player can have two roles in this system: the submitter, which submits a proposition to be verified by the system, and the verifier, which verifies the validity of each assigned proposition. 

\subsection{System Description}

\begin{figure}[t]
\centerline{\includegraphics[width=  \textwidth]{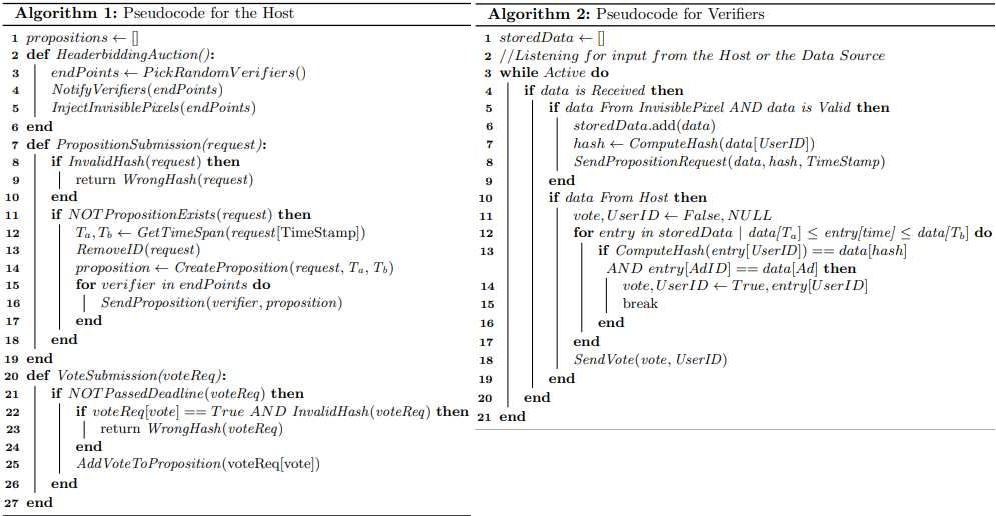}}
\centering
\caption{Pseudocode for the Host and Verifiers}
\label{algs}
\end{figure}

The system overview is shown in Figure ~\ref{fig2}, and the pseudocode for the host and verifiers is shown in Figure ~\ref{algs}. The description of each step is as follows:

\textbf{Step 1: }The process begins when a website visitor attempts to load a web page. Header bidding initiates before the body loads. During header bidding, platforms like Prebid collect visitors' data and hold an auction where DSPs compete for ad placements. Meanwhile, Prebid notifies the host of \toolad\ to inject some invisible pixels into each ad placeholder. These invisible pixels are placed in various random locations within the placeholder. This is because some discrepancies are often due to a faulty pixel, the user scrolling halfway through an ad (one pixel gets scrolled on and one does not), or connection issues from the pixel to the endpoint. By introducing multiple invisible pixels, we can reduce these issues.

\textbf{Step 2: } Unlike most Blockchain designs, all the nodes receive the input data; in \toolad, only a few randomly selected verifiers receive ad data. The reason is that the data is received through multiple invisible pixels, and adding too many pixels can decrease the website's performance. Also, millions of ad transactions occur per second, so transmitting all these transactions to every node can increase the cost and decrease the network's performance. Thus, once the pixels are injected, the host picks some verifiers randomly to be the endpoints of these pixels and receives ad data. One of the endpoints becomes the host itself. When the auction is over, the results get instantly relayed to  \toolad \ and other middlemen. Finally, the website gets loaded for the visitor.

\textbf{Steps 3 and 4: } When the website visitor engages with an ad, the JavaScript attached to the invisible pixels notifies the verifiers and host. The verifiers check if the data meets the criteria for a valid ad. If so, they submit a proposition request and pay the proposition price as a stake. A proposition request includes ad\_ID, user\_ID, Hash(user\_ID), Timestamp, and player\_ID. Hash(user\_ID) is the output of the cryptographic hash function for the user\_ID.

\textbf{Steps 5: }The host verifies the correctness of the request before generating a new proposition. The generated proposition by the host does not contain the user\_ID; instead, it contains the Hash(user\_ID). The user\_ID needs to be stored privately. Also, the Timestamp is converted to $T_a$ and $T_b$ with the Timestamp as their midpoint due to potential data delays. The proposition also contains the proposition deadline, the time by which the assigned verifiers must submit their votes. The verifier who submitted the proposition request becomes the submitter. The propositions are publicly available to all players, but only the assigned verifiers can vote.

\textbf{Step 6: } Assigned verifiers compute the hash of user\_IDs associated with the ad\_ID within the time frame of $T_a$ and $T_b$. If a hash matches the Hash(user\_ID) in the proposition, they vote True and submit the user\_ID as proof to the host; otherwise, they vote False. Failure to submit a vote by the proposition deadline results in a False vote, while submitting an incorrect user\_ID leads to a void vote. 

\textbf{Step 7: }The majority vote among the verifiers determines the proposition's outcome. Majority voters are rewarded, while the minority voters and those who submitted void votes are penalized. To resolve ties, the submitter's vote is counted as True when the owner requires an even number of verifiers for each proposition. Every time a player's proposition request is accepted or the verifier verifies a proposition, their proposition countdown (PCD) goes down by one. Every player initially receives the same value for PCD that the owner sets. Once the outcome is decided, all the nodes can access all the data transmitted since the system
is entirely transparent. This ensures the host or the transmission channels are not compromised.

\textbf{Step 8: }If a proposition is decided True, the associated ad impression counter increases and log-level data is logged. All data must be retained by both the host and each participant verifier for a predetermined duration before potential removal. Ad data has a finite lifespan, but there is the option to transfer it in bulk to a Blockchain network if needed.

\begin{table*}[!t]
\small
\newcommand{\tabincell}[2]{\begin{tabular}{@{}#1@{}}#2\end{tabular}}

\caption{Possible Outcomes of Propositions}
\label{table1}
\resizebox{\textwidth}{!}{
\begin{tabular}{|c|c|c|c|}
\hline
 \textbf{Majority} & \textbf{Verifiers with True Vote} &  \textbf{Verifiers with False Vote} & Submitter\\
\hline

\textbf{True} & +1 Point & -1 Point, - Proposition Price & +4 Points  \\
\hline

\textbf{False} & 0 Point &  0 Points &  -2 Point, - Proposition Price \\
\hline

\end{tabular}
}
\end{table*}

\subsection{Reward and Penalty}
Table~\ref{table1} summarizes each possible outcome. When a verifier submits a proposition request or votes True, the host checks the validity of Hash(user\_ID). If it is invalid, the verifier forfeits their stake and incurs a penalty of two negative points, discouraging potential Denial of Service (DoS) attacks.

If the majority of verifiers vote True on a proposition, the submitter is rewarded with four positive points, while True-voting verifiers earn one positive point each. However, False-voting verifiers lose their stake and receive one negative point each. Conversely, if the majority of verifiers vote False, the submitter incurs a penalty of two negative points and forfeits their stake, while all the other verifiers do not earn any points in this scenario. If a verifier votes True with the correct ID but remains in the minority, the wasted computational effort/cost is the penalty. The points get added to the player's points. If players do not lose their stake, they get it back, and the stake taken away from the dishonest players is added to the prize Pool. If the player point of a node becomes less than a threshold, they get banned from the network. The owner sets the threshold. 

Players receive a financial reward after their PCD gets down to zero. After this condition, if their player point is positive, they get a financial reward based on their points and the summation of all the players' positive points using the formula below. In this formula, $p_{v_i}$ is the player point of a given node, $u(V_j)$ is the step function, and $\sum_{j=1} p_{v_j} u(v_j)$ is the summation of the player points of all the Players with positive points. The point and PCD of each player get reset after they are paid. 

\[\text{Verifier Prize} = \text{Prize Pool} \cdot \frac{ p_{v_i}}{\sum_{j=1} p_{v_j}u(v_j)} \]

\subsection{Log-level Data}

\toolad's precise ad impression count helps resolve ad discrepancies among parties due to its cross-verifiable, transparent, and reliable nature. Unlike traditional methods, \toolad\ uses multiple independent endpoints, providing unbiased and comprehensive data for verification. Nonetheless, understanding the root causes of discrepancies often requires analyzing log-level data.

Log-level data that contains server logs provides detailed information about specific events like ad impressions, ad requests, and bid requests. This data contains essential attributes for each ad impression and helps parties understand the auction process, fees, involved parties, and visitor data. By leveraging this comprehensive data, involved parties can identify the technical errors in recording ad data and, hence, the root causes of ad statistics discrepancies \cite{loglevel}.

Many publishers face challenges in recording and storing log-level data due to its massive volume, complex infrastructure requirements, and high maintenance costs \cite{loglevel}. However, \toolad \ overcomes these limitations with its distributed and decentralized design. It only picks a few nodes per ad transaction, and the specific verifiers receive and store the data and provide it as needed at the end of each proposition verification period. Moreover, \toolad\ is designed only to record the data required in the verification process and find the causes of discrepancies; thus, it complies with the privacy policies as it only requires data similar to first-party cookies.

\subsection{Lazy Voting Prevention and Nash Equilibrium Incentive}

Lazy voting happens when players simply choose True or False without much thought or effort, realizing that either option is more likely to win. This boosts dishonest voting and can seriously affect the accuracy of the outcome. \toolad \ prevents lazy voting by requiring verifiers to provide proof when voting True. Meanwhile, not voting is considered as voting False, and lazy voting False is associated with more risk than reward. Also, submitting an invalid proposition request has financial penalties. Thus, submitters should not submit with no effort either.

In game theory, a Nash equilibrium (NE) is a scenario where no player gains an advantage by altering their strategy, provided all other players maintain their current strategies. In \toolad, at NE for the decision to what to vote, the best strategy is to vote True, but voting True requires proof. Having the proof suggests the verifier is honest and has calculated the reverse hash. In this case, the average payoff for each verifier is 1/2 points and no stake loss. False is only the best strategy if the proof is not found because voting False has an average payoff of -1/2 points and  1/2 stake loss. Even in that case, the honest behavior is voting False because the verifier has to attempt to calculate the reverse hash, and if it is not found, they know the correct vote is False. Therefore, if verifiers can find proof, voting True is the most optimal strategy, which is the honest behavior, and if they cannot, then voting False is the only option, which is also the honest behavior. In this case, False is the best option for all the nodes, making the majority vote False. Therefore, honest voting is the most optimal strategy.

\subsection{Limitations of \toolad}
\toolad\ aims to solve the ad impression discrepancy problem in header bidding, and it does not support other types of online advertising. Also, the system relies on accessing ad placeholders on a website to inject invisible pixels. Hence, permission from website owners is required. Finally, like most other decentralized systems, the quality of the network relies on the number of available nodes. Hence, multiple nodes are required to be active at all times.

\section{Experimental Evaluation}
\label{evaluation}
\subsection{Experimental Setup}
Multiple experiments were performed to analyze the feasibility of \toolad\ and the impact of some of the main system parameters on the accuracy of outcomes. The experiments were run on an Azure Standard E96ds v5 virtual machine equipped with 96 VCPUs and 672 GiB memory on Ubuntu 20.04 OS. Also, the latest versions (on July 2023) of nodejs, Docker, and PostgreSQL were installed. The source code and the results of the experiments can be found in \href{https://anonymous.4open.science/r/AdCahin-D274/}{here}\footnote{\url{https://anonymous.4open.science/r/AdCahin-D274/}}.

\subsection{Optimal Rate of Honesty and the Number of Verifiers}

\label{experimet1}

The experiment aims to demonstrate the effect of the honesty rate of each node and the number of verifiers per ad on the accuracy of \toolad. Honesty rate mimics the rate of accuracy or honesty of a node, as nodes may intentionally or unintentionally vote incorrectly for a proposition. We varied the honesty rate of each node from 0 to 100 in 5-unit increments. 0 \% means not honest and 100 \% means most honest. Then 50\% means the verifier misses 50\% of the assigned tasks. We adjusted the number of verifiers per proposition for each case from 1 to 20 in 1-unit increments. We generated and sent 1,000 simulated ad data each time, totaling 420,000 entries at a rate of 10 ad data/s. In the ideal scenario, all 1,000 propositions would be True, serving as the ground truth. Verifiers were randomly selected from a pool of 100 available nodes. Additional parameters included the player's initial balance (10,000), proposition price (1), proposition deadline (2 seconds), proposition delay (1 second), and threshold point (-5).

 \begin{figure*}[t]
\centerline{\includegraphics[width= \textwidth]{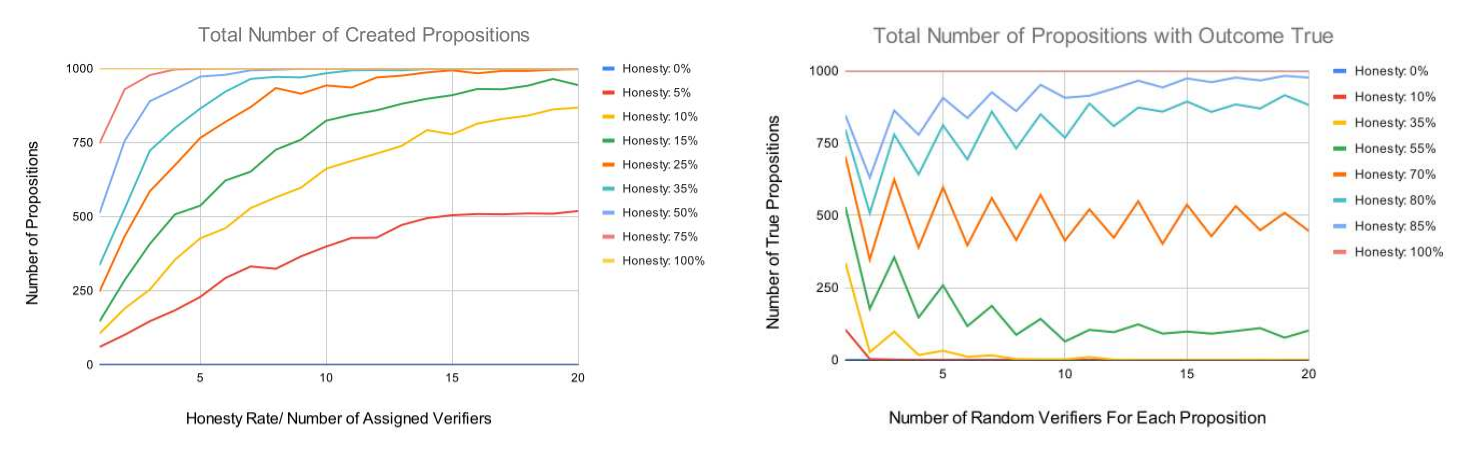}}
\centering
\caption{The effect of rate of honesty and the number of verifiers on the total number of propositions created (on the left) and propositions with outcome True (on the right). }
\label{ex1}
\end{figure*}

The results are shown in Figure~\ref{ex1} and are summarized for a more straightforward visual perspective. As shown on the left chart, as long as the honest rate is at least 35\% and there are at least 11 verifiers assigned to an ad, all the propositions are raised correctly. This is because if we have at least 11 verifiers, then at least 11 endpoints receive the ad data and can submit a proposition request. If the honesty rate of each of these verifiers is at least 35\%, then there is a high chance that for each ad transaction, at least one of them acts honestly. This shows how efficiently \toolad's mechanism captures the ad data. 

Moreover, as expected, the best and worst outcomes occur when all nodes have honesty rates of 100\% and 0\%, respectively, with all other results falling between them. An accuracy of 98\% is achievable for honesty rates of at least 85\%. However, there is a notable accuracy drop when the honesty rate decreases from 85\% to 80\% since as the honesty rate of all nodes decreases, the chance of the majority being on the inaccurate side increases. Therefore, it is essential to ensure that all nodes have an honesty rate of at least 85\%. This can be achieved by setting the proper system parameters and banning the dishonest nodes. 

For the honesty rate of 85\%, the best precision was achieved for 19 and 15 verifiers, leading to 980 and 983 True outcomes, respectively. Thus, the optimal number of verifiers per ad is recommended to be 15, balancing accuracy and resource allocation since each additional verifier increases the cost, as it necessitates an extra invisible pixel for each ad. This demonstrates \toolad\ is accurate and feasible for validating ad data and can decrease the discrepancy rate from up to 20\% to only 2\%.

 \begin{figure*}[t]
\centerline{\includegraphics[width= \textwidth]{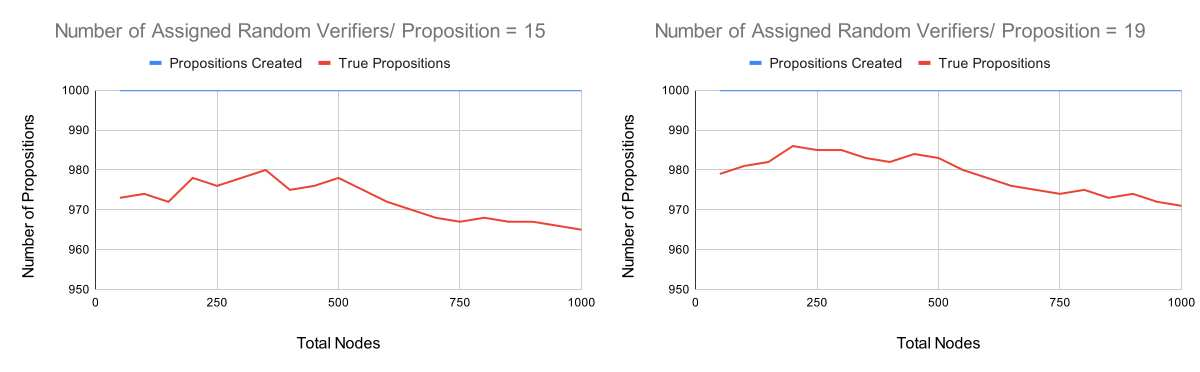}}
\centering
\caption{The effect of the total number of nodes on performance when 15 verifiers are chosen per proposition (left chart) and 19 verifiers are chosen (right chart). }
\label{ex2}
\end{figure*}

\subsection{Optimal Number of Nodes}
\label{experimet2}
This experiment aims to identify the optimal total number of nodes for \toolad. In this experiment, we varied the available nodes from 50 to 1,000 in 50-unit increments. For each case, we generated and sent 1,000 simulated ad data to 15 randomly selected verifiers with an honesty rate of 85\%, as determined in Experiment 1. We also repeated the experiment with 19 verifiers to compare the results. The rest of the parameters are similar to the ones in Experiment 1. 

Based on the results in Figure~\ref{ex2}, the number of nodes in the \toolad\ network has only a minor impact on system performance. This is because, in \toolad, only a few nodes receive each ad's data and can initiate proposition requests. Consequently, the data source sends data to a few selected endpoints out of the total nodes, and the network handles a few proposition requests. Therefore, increasing the number of nodes does not substantially affect the performance. 

In both cases, performance slightly drops when the number of nodes exceeds 550, but this decrease is insignificant. This slight drop is likely due to the resource limitation of the hardware we used during our tests. 

Generally, increasing the number of nodes decreases the chance of Sybil attacks and enhances accuracy, but it can increase the cost. However, in \toolad, the data is sent to a limited number of randomly selected nodes; increasing the number of nodes does not significantly increase cost but can improve system accuracy. Therefore, at least 200 active nodes are recommended at all times. 

Furthermore, selecting 15 verifiers instead of 19 drops the accuracy by only 0.5\%. Thus, the advantage of having four extra verifiers is insignificant, but it increases the costs. Thus, the optimal balance between accuracy and resource allocation is achieved with 15 verifiers from at least 200 nodes.

Our analysis shows that every extra 50 nodes cost 0.4 to 0.8 \% CPU and RAM usage and no noticeable disk usage. Also, On average, the latency (neglecting network delays) is about 89 ms, indicating that \toolad\ can scale effectively. This is because the data size is small, and the consensus protocol is efficient in time, cost, and resource usage. This shows that  \toolad\ can scale to the current transaction rate for online advertisement. This observation is based on limited resources, with a single machine handling all tasks. Better performance is expected in a real-world scenario where each node has its own machine.

\subsection{Sybil Attack Success Rate}
\label{experimet3}
This experiment aims to simulate a Sybil attack on \toolad. Commonly, in a successful Sybil attack, an entity gains control of 51\% of the network, preventing proposition creation or manipulating proposition outcomes. In this experiment, we selected fifteen verifiers from a pool of one thousand nodes. The attack's success rate is determined by the corrupted nodes' ability to prevent proposition creation or change the proposition outcomes to False.

We varied the rate of the corrupted nodes from 0\% to 100\% in 5-unit increments of the total nodes. If a node is corrupted, its honesty rate becomes 0\%. Also, we decided to have two cases with the honesty rate of uncorrupted nodes as 85\% and 100\%. Then, we repeated the experiment with 200 total nodes.

\begin{figure*}[t]
\centerline{\includegraphics[width=  \textwidth]{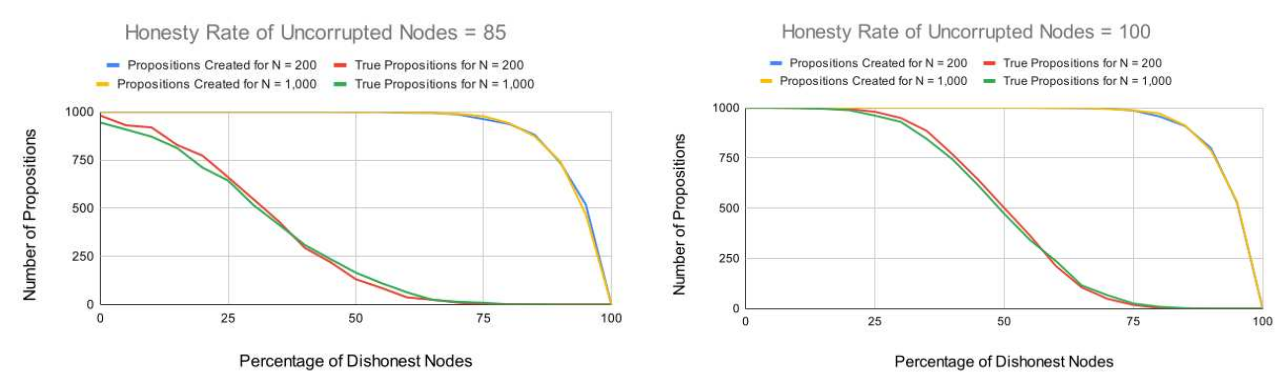}}
\centering
\caption{Sybil attack success rate when the honesty rate of uncorrupted nodes is set to 85 (left chart) and 100 (right chart). N represents the total number of nodes.}
\label{ex3}
\end{figure*}

As depicted in Figure~\ref{ex3}, the results show a similar trend for both the 200-node and 1,000-node cases. When uncorrupted nodes had a 100\% honesty rate, even with 50\% of the network being corrupted, about half of the propositions remained accurate. An entirely successful Sybil attack required the corrupted nodes to control at least 80\% of the network. However, reducing the honesty rate of uncorrupted nodes to 85\% increased the attack's success rate. In this case, gaining 70\% of the network was enough to launch a successful Sybil attack. 

These results represent an improvement over typical scenarios where attackers only need 51\% of the network for a successful attack. Also, since increasing the total number of nodes has an insignificant effect on the overall performance of \toolad, we can acquire more nodes to make the Sybil attack more expensive.

\textbf{Sybil Attack Prevention:} In \toolad, Sybil attacks are very costly because only a few nodes can access an ad transaction and raise or verify a proposition. An adversary must gain more than 75\% of the entire network to perform a successful attack. Also, the adjustable parameters and the point and stake system make such an attack more complicated, riskier, and more expensive. Finally, an authentication step can be added during sign-up if needed.

Mathematically, if $n$ verifiers are picked out of $N$ nodes while $j$ nodes are dishonest, the probability of at least 51\% of the verifiers being dishonest is shown below. Plugging in the suggested parameters confirms the findings in Figure ~\ref{ex3}. 

\begin{equation}
  P[\text{Sybil\_attack\_success}] = 
  \sum\limits_{i=\lfloor n/2 \rfloor + 1}^{n} 
  \frac{\binom{j}{i} \binom{N - j}{n - i}}{\binom{N}{n}}
\end{equation}

\section{Economic Model}

In this cost analysis, we make the following assumptions:

\begin{enumerate}

    \item Each invisible pixel sends at most 1 KB of data per ad transaction to each of the verifiers. For simplicity,  $n = 1$KB where $n$ is the size of each ad data.

    \item Since the computational cost is low (calculating a few hashes per proposition), the verifiers' bottleneck for the number of verified ads is the internet bandwidth of the verifiers. 
    
    \item Each verifier is willing to use at least 10\% of their internet bandwidth on receiving ad data and sending propositions and votes. 

    \item Each ad data is sent to 15 verifiers as shown in Experiment 1 (Section ~\ref{experimet1}). Thus, $V = 15$, where the $V$ is the number of randomly assigned verifiers per ad transaction.
    
    \item The publishers are willing to allocate 1\% of their annual revenue as the reward of the verifiers. 
\end{enumerate}

\textbf{Revenue:}
The average Internet Speed for downloading is 70 Mbps (Fixed Broadband and cellular) worldwide\footnote{\url{https://www.speedtest.net/global-index}}. Considering $n = 1 $Kb and assuming that each verifier is willing to use at least 10\%  of their internet bandwidth, each active verifier can receive ad data at 7,000 ads per second. 

There are 300 billion global ad impressions daily. This equates to about 3.5 million transactions per second. Therefore to estimate $N$, the minimum number of active nodes required at all times to be able to verify all the worldwide ad transactions, we can use the following formula:

\[N =  \frac{ 3.5 \cdot 10^6 \cdot V}{7 \cdot 10^3} = 500V\] 

Hence, since we assumed V = 15, we need at least 7,500 verifiers. However, as indicated in Experiments 2 and 3 (Sections ~\ref{experimet2} and ~\ref{experimet3}), it is best to randomly pick these 15 verifiers from a pool of at least 200 active nodes. Hence, the minimum number of active nodes at all times becomes 100,000. This is, of course, to verify the header bidding ad transaction of the entire world. 

The revenue of header bidding was about \$414B in 2022 \cite{MarketSize} \cite{HeaderBidding}. Thus, assuming 1\% can be allocated to the prize pool, the annual prize pool is at least \$4B. This means each active node can make about \$40K per year on average. We assume the other \$0.14B plus the stake taken from the dishonest nodes goes toward the system's host. 

\textbf{Cost:} Each node needs constant access to the Internet with an average speed costs about \$380 per year on average worldwide\footnote{\url{https://www.numbeo.com/cost-of-living/country_price_rankings?itemId=33}}. 

The other costs are associated with the computation, storing, and transmitting data. Since the ad data is only sent to the selected verifiers, each verifier receives 7,000 ad data per second, and each ad data is 1KB. Each verifier needs to be able to process and store 7 Mb/s of data. Most of the computational cost goes towards calculating SHA3, and most modern computers can calculate it at a 7 Mb/s rate. For instance, a machine like Azure B16ms VM would cost around \$4,000 per year.

Regarding the storage needed, 7,000 ad transactions per second, which contains 1Kb of data, each equates to about 18 TB per month. Each ad data has an average life span of 1 month. After a month, the DSPs pay the SSPs and publisher, and the data can be deleted. Hence, 18 TB per month is enough for each node. In this case, an Azure P70 storage quoted around \$18.7K per year can be used. 

\textbf{Profit:} On average, the total revenue of each node is \$40K while the overall cost is around \$23K annually. Hence, the average profit for a node is \$17K/year. 

To compare this number to Bitcoin mining, there are about 1 million Bitcoin miners worldwide, and 330K Bitcoin gets mined every year \cite{bitdata}. Therefore, on average, each Bitcoin miner mines 0.33 Bitcoins per year. The current price of Bitcoin is about \$63K, making the average miner's revenue equal to \$21K/year. Meanwhile, the average cost of mining one Bitcoin is around \$26.5K, making an average Bitcoin miner's cost around \$9K/year \cite{Bitcost}, \cite{Bitcost2}. Thus, on average, each Bitcoin miner profits about \$12K/year \cite{Bitcost3}. 

As a result, on average, a node on \toolad\ can make \$17K/year, comparable to a Bitcoin miner's profit. Additionally, Bitcoin mining requires high-end equipment with a challenging setup. Also, Bitcoin mining has high energy consumption, which has raised many concerns. Finally, Bitcoin supply is projected to end by 2028 \cite{bitdata} while online advertising demand keeps growing\cite{hpaper}. Therefore, there are many advantages for nodes on \toolad\ network. 

\section{Conclusions}
\label{Conclusion}
Ad discrepancy fraud is a significant problem in the online advertising industry, causing substantial financial losses. The involvement of numerous intermediaries, a lack of trust and transparency, inadequate regulations, and a complex supply chain exacerbate this problem. To address this issue, we present \toolad, a decentralized, distributed, and verifiable oracle solution to header bidding. \toolad \ detects and prevents online advertisement fraud by acquiring multiple independent agents that receive and record log-level data and a consensus protocol that determines the validity of each ad data. One of the highlights of this design is that, unlike the other decentralized designs, it is compatible with the current infrastructure and does not replace or remove any current major parties. This paper shows that \toolad is easy to adapt and implement, scalable, accurate, and secure. As a result, it can improve the current header bidding mechanism and decrease the rate of discrepancies to 2\%.  

In the future, we plan to expand our work to support other advertising platforms in addition to Prebid and conduct practical testing with the assistance of some SSPs.

\newpage

%
%
%
%

\bibliographystyle{splncs04}
\Urlmuskip=0mu plus 1mu
\bibliography{Ref}

\end{document}